\documentclass[aps, prl, amsmath, floats,floatfix, twocolumn, superscriptaddress, nofootinbib, showpacs]{revtex4}

\usepackage{amssymb}
\usepackage{amsmath}
\usepackage{verbatim}
\usepackage{mathrsfs}
\usepackage{amsfonts}
\usepackage{latexsym}
\usepackage{epsfig}
\usepackage{color}
\usepackage{graphicx,subfigure}
\usepackage{units}
\usepackage{dcolumn}
\usepackage{bm}
\usepackage{multirow}

\newcommand{\lsim}{\mbox{\raisebox{-.9ex}{~$\stackrel{\mbox{$<$}}{\sim}$~}}}

\def\thebiblio#1{
\begin{center}\bf \large References
\end{center}
\list
{[\arabic{enumi}]}{\settowidth\labelwidth{#1.}\leftmargin\labelwidth
 \advance\leftmargin\labelsep
 \usecounter{enumi}}
 \def\newblock{\hskip .11em plus .33em minus -.07em}
 \sloppy
 \sfcode`\.=1000\relax}

\begin{document}

\preprint{}
\title{%
Analysis of the Very Inner Milky Way Dark Matter Distribution \\and Gamma-Ray Signals }

\author{V. Gammaldi}
\affiliation{Department of Astrophysics, SISSA, Trieste, Italy}
\affiliation {Istituto Nazionale Fisica Nucleare INFN}
\affiliation {Universidad Complutense Madrid, Madrid, Spain}
\affiliation {Instituto de Astronom\'ia, Universidad Nacional Aut\'onoma de M\'exico, A.P. 70-264, 04510 CDMX, Mexico}
\author{V. Avila-Reese}
\affiliation {Instituto de Astronom\'ia, Universidad Nacional Aut\'onoma de M\'exico, A.P. 70-264, 04510 CDMX, Mexico}
\author{O. Valenzuela}
\affiliation {Instituto de Astronom\'ia, Universidad Nacional Aut\'onoma de M\'exico, A.P. 70-264, 04510 CDMX, Mexico}
\author{A. X. Gonzalez-Morales}
\affiliation{ CONACYT- Departamento de F\'isica, DCI, Campus Le\'on, Universidad de
Guanajuato, 37150, Le\'on, Guanajuato, M\'exico.}

\date{\today}
\begin{abstract}
  We analyze the possibility that the HESS $\gamma$-ray source at the Galactic Center could be explained as the secondary flux produced by annihilation of TeV Dark Matter (TeVDM) particles with locally enhanced density, in a region spatially compatible with the HESS observations themselves.
We study the inner 100 pc considering  (i) the extrapolation of several density profiles from
state-of-the-art N-body + Hydrodynamics simulations of Milky Way-like galaxies, (ii) the DM spike induced by the black hole, and (iii) the DM particles scattering off by bulge stars. We show that in some cases the DM spike may provide the enhancement in the flux required to explain the cut-off in the HESS J1745-290 $\gamma$-ray spectra as TeVDM. In other cases, it may helps to describe the spatial tail reported by HESS II at angular scales $\lesssim 0.54^\circ$ towards Sgr A$^*$. 
\end{abstract}
\maketitle

Observations of  High Energy (HE) and Very High Energy (VHE) $\gamma$-rays  from the Galactic Center (GC)
have been reported by different collaborations such as CANGAROO \cite{CANG}, VERITAS \cite{VER},
HESS  \cite{Aha, HESS}, MAGIC \cite{MAG} and Fermi-LAT  \cite{Vitale, ferm}. However, the uncertainty  associated to the  Dark Matter (DM) distribution at the GC
 affects the possible explanation of observed signals products from DM-SM interactions rather than astrophysical sources \cite{DMrev, CaloreBg}. 
New observations of the GC region were made with the 5-telescope HESS II array \cite{HESS2015}. The data confirms an excess source at the position of the super massive Black Hole (BH) Sgr A$^*$ at a significance of 40$\sigma$ in the $0.015$ deg$^2$ and a relatively long tail to $0.3$ deg$^2$.  Such a tail indicates the contribution of diffuse $\gamma$-ray emission at large distance from the source. These features may be an indication of either some extension of the emission previously seen by HESS, or escape of emission outside the excluded region due to the reduced angular resolution \cite{HESS2015}. However, the origin of the cut-off around 30 TeV in the inner 10 pc at the GC  \cite{HESS} is yet an open question. In \cite{Gammaldi} it was shown that such a cut-off in the HESS J1745-290 $\gamma$-ray spectra is well fitted by secondary emission from annihilation of thermal TeVDM particles with a background component that appears compatible with the lower energy FERMI-LAT data from the same region. However, for the commonly used NFW \cite{Navarro:1996gj} halo mass distribution, a boost factor of $\approx 10^3$ was required.
Keeping the hypothesis of thermal TeVDM candidate, some effects that can contribute to the needed enhancement are the inner halo contraction due to baryons, the presence of DM clumps in the Galaxy, and a DM-spike at the GC. The latter might be originated by the presence of the BH Sgr A$^*$. This possibility has been investigated in the literature, both in the classic and relativistic approach \cite{BHgrowth,BHrel}. The resultant boost factor could be of several orders of magnitude depending on the assumptions about the initial density profile and BH growing history\cite{BHgrowth,Alma}. Further studies have taken into account the dynamical effect of the stars \cite{stars} or the instantaneous, or slow growth, of the BH \cite{BHgrowth} contributions that effectively reduce such enhancement.

\begin{figure}[h!]
\begin{center}
\epsfxsize=13cm
\resizebox{7.8cm}{5.6cm}
{\includegraphics{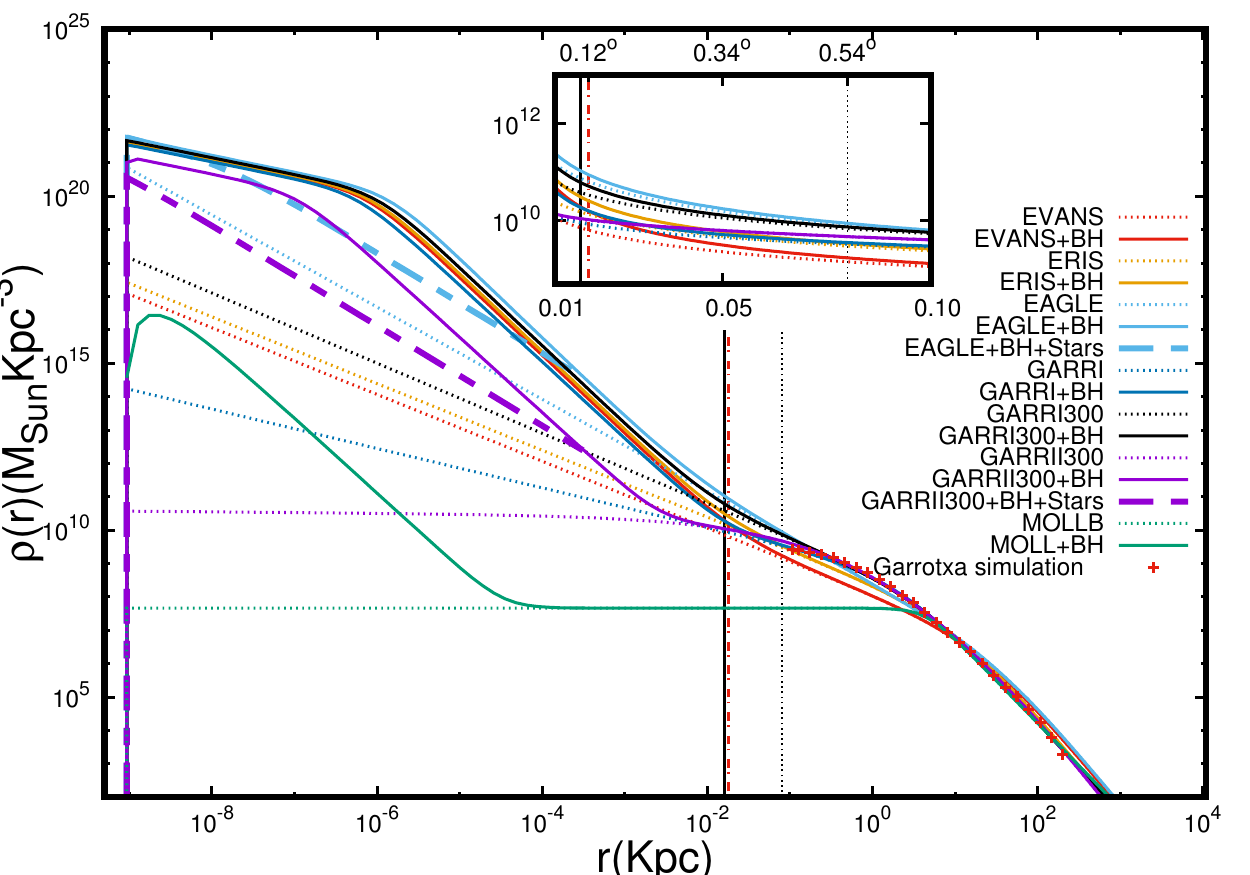}}
\caption {\footnotesize{ DM-halo profile from different simulations extrapolated down to the GC (dotted lines), and their BH-related spike (full lines). The effect of the stars in the inner 0.34 pc (dashed lines) is shown for two profiles: GARR-II300 and EAGLE (violet and grey dashed line). The Garrotxa simulation is used as reference (red points). Vertical lines indicates the region observed by HESS with a $0.12^\circ$ and $0.54^\circ$ resolution angle, respectively. The morphology of the DM density distribution at such angular resolution is shown in detail in the inset. (colours online)
}}
\label{BH}
\end{center}
\end{figure}

In this paper, we performed a in-depth study of the astrophysical factor for a $\gamma$-ray flux generated from a BH DM-spike at the GC, with respect to several initial underlying DM-halo distributions, accounting also for the effect of the stars. Then, we study the radial and angular dimension of such a spike and its compatibility with the spatial morphology of the HESS $\gamma$-ray signal.\\

The differential $\gamma$-ray flux from annihilation of DM particles reads:
\begin{equation}
\frac{d \Phi_{\text{DM}}}{dE} =\sum^{\text{channels}}_i \frac{\langle\sigma v\rangle_i}{2} \cdot \frac{dN_{i}}{dE}
 \cdot \frac{{\Delta\Omega\,\langle J \rangle}_{\Delta\Omega}}{4 \pi m_\text{DM}^2}\,,
\label{eq:totalflux}
\end{equation}
where $\langle \sigma v\rangle_i$ is the annihilation cross section in each SM channel $i$, $dN_i/dE$ is the $\gamma$-ray spectrum produced by subsequent hadronization or decay events of SM particles, $m_\text{DM}$ is the mass of the DM particle and $\Delta\Omega$ is the solid angle. The astrophysical factor $\langle J \rangle_{\Delta\Omega}$ accounts for the DM density distribution $\rho(r)$ in the source, and it is given by the integration along the line of sight (l.o.s) from the source to the observer, $l(\theta)$: 
\begin{eqnarray}
\langle J \rangle_{\Delta\Omega}= \frac{1}{\Delta\Omega}\int_{\Delta\Omega}\text{d}\Omega\int_0^{l(\hat\theta)_{max}} \rho^2 [r(l)] dl(\theta)\,,
\label{flux}
\end{eqnarray}
where, $r^2 = l^2 + D_\odot^2 -2D_\odot l \cos \theta$  and  $D_\odot \simeq 8.5$ kpc are the  distance  from the GC  to any point of the halo, and to the Sun respectively. The maximum distance for integration is given by the the edge of the DM distribution in the $\theta$  direction, $l(\hat\theta)_{\rm max} = D_\odot \cos \hat\theta + \sqrt{r^2-D_\odot^2 \sin \hat\theta}$. Notice here that  $\hat\theta$ is related with the morphology of the source and the telescope angular resolution.  The photon flux is maximized in the direction of the GC, and must be averaged over the solid angle of the detector.
For detectors with sensitivities in the TeV regime, the solid angle is typically of order $\Delta \Omega = 2 \pi ( 1 - \cos \hat\theta ) \simeq 10^{-5}$ or $\hat\theta\approx 0.1^\circ$, as it is the case for the HESS Cherenkov telescopes array.

\begin{center}
\begin{table*}[ht!]
\begin{center}
\resizebox{\textwidth}{!}
{
\begin{tabular}[b]{|c|c|c|c|c|c|c|c|c|c|}
\hline
\hline
Profile & $\rho_s\left(\text{M}_\odot/\text{Kpc}^{-3}\right)$ & $r_s\left(\text{Kpc}\right)$ & $r_\text{vir} $ (kpc) & $\gamma$ & $\alpha$ & $\beta$ & $\rho_\odot(\text{GeV}\text{cm}^{-3})$ & $\text{R}_\text{sp}$ (pc) & $\theta_\text{sp}^\circ(\text{deg})$  \\
\hline
EVANS & $5.38\times10^6$ & $21.5$ & $215$ & 1 & 1 & 3 & 0.27 & 24 & 0.16\\
\hline
GARR-I & $4.97\times10^8$ & $2.3$ & $230$ & $0.59$ & $1$ & $2.70$& 0.33 & 16 & 0.11\\
\hline
GARR-I300 & $1.01\times10^8$ & $4.6$ & $230$ & $1.05$ & $1$ & $2.79$& 0.33 & 11 & 0.07\\
\hline
GARR-II300 & $2.40\times10^{10}$ & $2.5$ & $230$ & $0.02$ & $0.42$ & $3.39$& 0.34 & 2.3 & 0.01\\
\hline
ERIS & $2.25\times 10^7 $  & 10.9 & 239 &  1 & 1 & 3 & 0.35 & 16 & 0.11 \\
\hline
MOLL & $4.57\times 10^{7}$  & $4.4$ & 234 & $\sim 0$ & $2.89$ & $2.54$  & 0.29 & 0.034 & $0.0002$\\
\hline
EAGLE & $2.18\times10^{6}$  & $31.2$ & 239 & $1.38$& $1$ & $3$& 0.31 & 6.4 & 0.04\\
\hline
\end{tabular}
}
\end{center}
\caption{ \footnotesize{Parameters of different DM density profiles as in Eq. (\ref{Eqrho}). We compare the EVANS DM-only simulation with the Hydro simulations GARR, ERIS, MOLL, and EAGLE.  See text for the different fits for GARR. We also provide the virial radius $r_\text{vir}$, the corresponding local DM density $\rho_\odot$, the linear dimension of the DM spike, $R_\text{sp}$, and its projected angular dimension on the sky, $\theta_\text{sp}$. }}
\label{tab:Rspike}
\end{table*}
\end{center}

{\bf The DM-halo profile.-}
The Galaxy's DM-halo density profile is one of the main sources of uncertainty in the $\gamma$-ray flux estimates. It is only well constrained by observations at scales above $\sim 5$ kpc \cite{Iocco}. At smaller scales one uses either: extrapolations of the density profile matching at larger scales or the results from cosmological simulations. 
We use density profiles from state-of-the-art N-body + Hydrodynamics simulations of MW-like galaxies in the $\Lambda$CDM cosmology: Mollitor et al. (MOLL, Halo B; \cite{MOLL}), ERIS \cite {ERIS},  Schaller et al. (EAGLE-APOSTLE, Halo 1; \cite{EAGLE}), and Garrotxa (GARR; \cite{GARR}).
For comparison to \cite{Gammaldi}, we include the only-DM simulation used by them (EVANS, \cite{EVANS}). At intermedium and large radii, the DM halos in all the Hydro simulations are described roughly by the NFW profile. 
At radii $\lesssim 5-10$ kpc (depending on the simulation), a bump with respect to the NFW profile is observed, likely due to halo gravitational contraction after baryons condensation \cite{Blumenthal}. At the innermost radii, the profiles tend to flatten, probably due to the effects of supernova feedback \cite{SNeffects}. A general function to describe the profiles is:
\begin{equation} 
\rho_h\left(r\right)=\frac{\rho_s}{\left(\frac{r}{r_s}\right)^{\gamma}\left(1+\left(\frac{r}{r_s}\right)^\alpha\right)^{\frac{\beta-\gamma}{\alpha}}}\, \\
\label{Eqrho}
\end{equation}
where $\rho_s$ is the normalization parameter and $r_s$ is the scale radius. The NFW profile corresponds to $\alpha,\beta,\gamma=(1,3,1)$. In Table \ref{tab:Rspike} we report the fit parameters for the different simulations mentioned above. For ERIS (softening length of 120 pc), the authors provide the fit to a NFW profile, though this function does not describe well the inner regions. However, the reported NFW fit implies a halo much more concentrated than in the only-DM simulation, with a value of $\rho_{\odot}$ compatible with observations. When $\gamma$ is left as free parameter, its determination depends on the minimum radius assumed in the fit, $r_{\rm min}$. For GARR, we fit Eq. (\ref{Eqrho}) as done also in \cite{MOLL} for MOLL. The spatial resolution limit in GARR and MOLL (cell length at the maximum level of refinement for this kind of AMR simulations) are $109$ and $150$ pc, respectively. According to convergence tests, a more suitable value of the spatial resolution seems to be $2-3\times$ the cell length, so for GARR we probe both $r_{\rm min}=109$ and 300 pc (GARR-II300 fit for the latter). The flattening of the density profile in GARR starts at $\sim 500$ pc,
but the slope with $\gamma_{}\approx 0$ is attained only at $r<r_{\rm min}$. In MOLL, a flat, $\gamma_{}\sim 0$ core is seen from $\sim 3$ kpc, a radius much larger than $r_{\rm min}$. We have found that five free parameters do not improve the statistical significance of the fit of GARR data with respect to four free parameters ($\alpha$ fixed to 1). Besides, in the latter case, $\gamma$ describes better the measured slope at $r_{\rm min}$. Therefore, Eq. (\ref{Eqrho}) with $\alpha=1$ is a more adequate fitting function (GARR-I and GARR-I300 fits). For EAGLE the fit of Eq. (\ref{Eqrho}) si performed with $\alpha=1$ and $\beta=3$ fixed and from the convergence radius (559 pc; the softening length is 132 pc)\cite{EAGLE}. As in GARR, a flattening of the inner profile from $\sim 2\times$ the convergence radius is also seen in this simulation. In fact, the authors show that the extrapolated slope should be shallower than the given by the fit in order to conserve the enclosed mass (see \cite{EAGLE} for details). However, it is not clear at which inner radius the slope should flatten. The extrapolations of the DM density profiles presented in Table \ref{tab:Rspike} cover a range of possibilities from cuspy to flat cores at the radii of interest ($\sim 100$ pc); at smaller radii the BH-induced DM spike starts to dominate (see below and Fig. \ref{BH}). \\
\\
 {\bf The inner DM slope.-} We now consider how the extrapolated density profiles presented in Tab. \ref{tab:Rspike} would be modified considering the adiabatic growth of the BH at the GC \cite{GS}. The DM-density appears locally enhanced in a region of radius $R_\text{sp}=\alpha_\gamma r_s(\text{M}_\text{BH}/\rho_s r_s)^{1/(3-\gamma)}$, where $r_s$, $\rho_s$ and $\gamma$ are defined in Eq. (\ref{Eqrho}) and $\alpha_\gamma$ is given in \cite{GS} for different profiles. In Tab. \ref{tab:Rspike} we give the value of $R_\text{sp}$ for each simulation. The only additions to the recipe outlined in \cite{GS} are that (i) we consider a BH mass $\text{M}_\text{BH}=4.5\times10^6\,\text{M}_\odot$ \cite{BHmass} and a correction factor to account for relativistic effects on the redistribution of DM around the BH \cite{BHrel} (as a consequence, the DM distribution vanish at  $r < 2R_{s}$ instead
of $4R_{s}$ as in \cite{GS}, where $R_s$ is the Schwarschild radius), and  (ii) we relax the assumption of circular orbits so that efficient annihilations at vey inner radii leads to a mild cusp instead of a DM plateau \cite{Diffusion} (for our assumption of isotropic velocity dispersions, the cusp goes as $r^{-1/2}$).  For our analysis,  we set the BH accretion time $t_\text{BH}\simeq 10$ Gyr and we consider a DM candidate with mass $m_\text{DM}\simeq50$ TeV and thermal annihilation cross section $\left<\sigma v\right>=3\times10^{-26}\text{cm}^3\text{s}^{-1}$. As mentioned above, this DM particle provides a good fit to the TeV cut-off detected by HESS in the $\gamma$-rays flux at the GC \cite{Gammaldi}. The upper limit in the DM-spike density depends on the DM candidate as $\rho_\text{ann}=m_\text{DM}/\left<\sigma v\right>t_\text{BH}$ at $r\lsim 10^{-3}$ pc. However, this effect is negligible when the profile is integrated between 0 and $l_\text{max}\approx r_\text{vir}$ on the l.o.s. (see Tab.  \ref{tab:Rspike} and Fig. \ref{BH}).  For completeness, we also consider the effect on the cusp of the scattering off of DM particles by bulge stars \cite{stars}. Because the GC is dynamically old, the affected innermost region ends up with a nearly universal density profile, independent of the initial DM density profile \cite{stars}. According to this work, the universal slope is $-3/2$ from a very inner radius (0.34 pc)\cite{034}. Further away, the slope tends to match the initial halo + spike profile. We use this dependence to construct the final profile accounting for the effect of the BH and the stars. \\

\begin{figure}[t]
\begin{center}
\epsfxsize=13cm
\resizebox{7.8cm}{5.6cm}
{\includegraphics{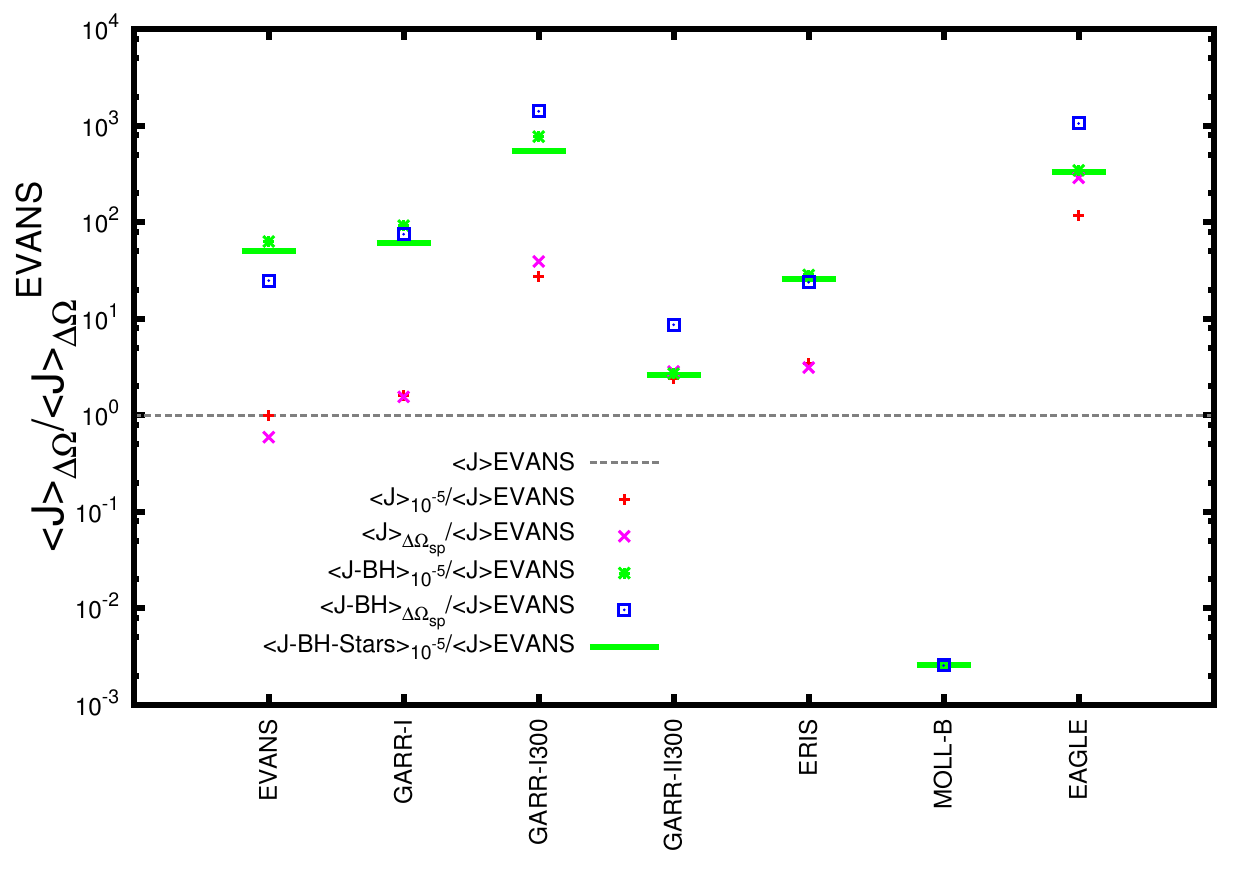}}
\caption {\footnotesize{Boost factors with respect to $\langle J \rangle_{\Delta\Omega}^{\text{EVANS}}$ for different DM profiles. The variation in the resolution angle, $\approx 0.1^\circ$($10^{-5}$ sr) (red/green symbols) and $\theta_\text{sp}$ (violet/blue symbol), does not (does) affect significantly  the $J$-factor in the absence (presence) of the BH spike. Horizontals lines indicates the variation when taking into account the effect of stars. The $J$-factor given by \cite{EVANS} and used in \cite{Gammaldi} is a lower limit in comparison with the DM density distributions from simulations with baryons, excepting for the cored distribution of MOLL. (colours online)}}
\label{Jboost}
\end{center}
\end{figure}
 
\begin{figure*}[]
\begin{center}
{\includegraphics[width=0.8\textwidth]{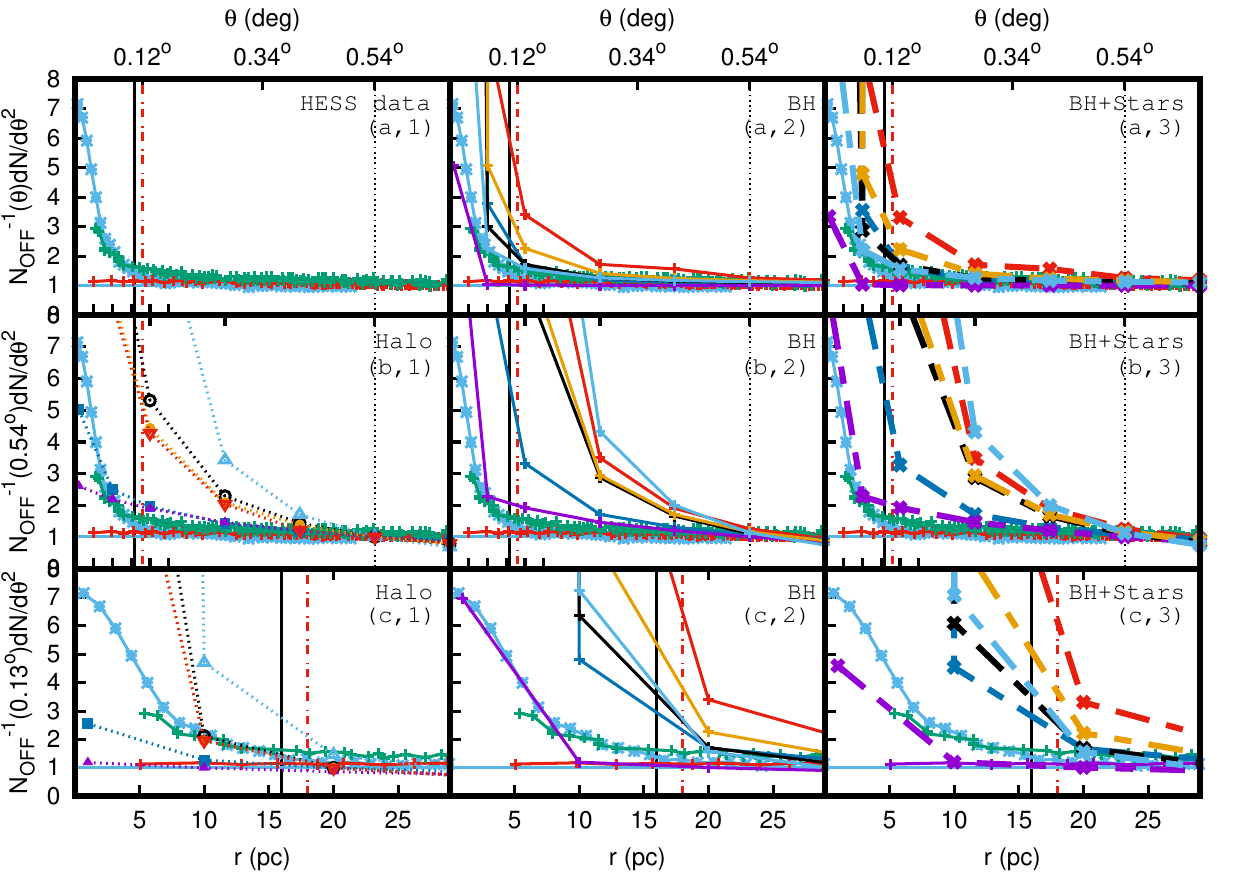}}
\caption {\footnotesize{HESS (blue points) and HESS II ON (green points) and OFF (red points) data. Three different background normalizations (Lines a, b, c) allow to compare three different cases for the ON source number of events, DM-halo (Column 1), BH-spike (Column 2) and stars effect (Column 3), to the OFF source background level. See text for details. Colours code as in Fig. 1. (colours online)}}
\label{HESSIvsII}
\end{center}
\end{figure*}

{\bf The astrophysical factor.-}
In Fig. \ref{Jboost} we present the J-factors calculated for different cases and normalized to the average factor obtained for the EVANS profile ($\langle J \rangle_{\Delta\Omega}^{\text{EVANS}}=280\times10^{23}\left(\text{GeV}^2\text{cm}^{-5}\text{sr}^{-1}\right)$) as was assumed in \cite{Gammaldi} for the analysis of the energy spectra in Eq. (\ref{eq:totalflux}) and TeV cut-off in the HESS data. In that figure, {\it{(i)}} The red crosses are for the extrapolated DM density profiles. The inner halo contraction due to baryons produces a boost factor (with respect to EVANS) of $\approx 20$ and 100 for GARR-I300 and EAGLE, respectively. {\it{(ii)}} The green and blue squares are for the DM profiles including the spike produced by the adiabatic growth of the BH, for the angles  $\approx 0.1^\circ$ and $\theta_\text{sp}=\text{ArcSin}[R_\text{sp}/D_\odot]$ (see Tab. \ref{tab:Rspike}), respectively. In this case, the  boost factors increase even more. For the GARR-I300 profile, this factor at  $\approx 0.1^\circ$($10^{-5}$ sr) attains a value of $\approx 500$ while for $\theta_\text{sp}$($\Delta\Omega_\text{sp}$), the factor is $>10^3$. Our results show that the boost factor can be maximized when the resolution angle of the telescope reaches the dimension of the spike. In fact, unlike the solid angle $\Delta\Omega$, which is independent from the source distance, the subtended angle depends on $D_\odot$.  As the underlying halo profile is more flattened, $R_\text{sp}$ and $\theta_\text{sp}$ are smaller; the extreme case is for the MOLL cored halo profile, where the spike appears at $\sim 0.001^{\circ}$, much lower than the current observational resolution. {\it{(iii)}} Finally, we show as green horizontal bars, the boost factors for the case of the angle at $10^{-5}$ sr, when the dynamical effect of stars on the DM-spike is taken into account \cite{stars}. This effect lowers the innermost cusp in a way that little depends on the underlying DM profile \cite{stars}, and it does not modify significantly the $J$-factor.\\
\\
{\bf The VHE $\gamma$-rays tail-}
We now focus on the spatial dimension and slope of the  DM-spike. As shown in Tab. \ref{tab:Rspike} and Fig. \ref{BH}, the extent of a BH-induced DM-spike is expected to be $\approx 1-20$ pc (depending on the DM-halo profile), and can be barely resolved with an angular resolution of $\approx 0.1^\circ$ that is typical of VHE $\gamma$-ray telescope such as HESS. For our spatial morphological analyses, we consider the angular distribution of the excess events in the two collections of HESS data in 2009 and 2015 (see Panel (a,1) Fig. \ref{HESSIvsII}). These data are the result of a deconvolution process of the signal with the Point Spread Function (PSF) of the instrument. Such a deconvolution process allows to extrapolate informations on the morphology of the source for angular resolution better than the nominal PSF \cite{BgPSF}. We consider the inner enhancement in the DM density distribution at the GC as the signal (ON source) above the DM-halo profile (OFF source) (in the sense of local enhancement), both of them integrated along the l.o.s.. Here, we hypothesize that, without such an enhancement, the level of secondary $\gamma$-rays produced by TeVDM annihilation in the halo remains undetected under the astrophysical background level. We normalize both the background levels (OFF source) to one, while the spatial morphology of the tail (ON source) is kept. Such an approximation allows to normalize both the data and the model to one, neglecting any factor of proportionality that is expected to be related to the particle physics part of Eq. (\ref{eq:totalflux}) and the data analysis. However, the background modelization in the ON source strongly depends on the data analysis. Generally speaking, it is an extrapolation of what is estimated to be the background in an OFF source that is around the ON source. Such an extrapolation can be performed with different methods depending on the kind of source and the adopted data analysis \cite{Bg}. For the 2009 HESS data, the combined Hillas/Model analysis  is developed as function of the resolution angle $\theta$ \cite{HESS}. On the other hand, for the 2015 HESS II data the level of the background contamination is estimated with the reflected region method \cite{HESS2015, Bg}.
These two factors (PSF deconvolution and background rejection) might strongly affect the shape of the spatial tail in the $\gamma$-ray signal. For such a reason, we adopt a first approximation in which we normalize the number of events $N_\text{ON}$ to the external $N_{\text{OFF}}(\theta\approx0.5^\circ)=100$ and 400 for \cite{HESS} and \cite{HESS2015} respectively, and we compare them with an also normalized DM distribution model: 
\begin{eqnarray}
\frac{1}{N_{\text{OFF}}}\frac{dN(\theta)_{\text{ON}}}{d\theta}&\propto&\frac{d \Phi_{\text{DM-spike}}(\theta)}{d \Phi_{\text{DM-halo}}(\theta)}=\nonumber\\
&= & \frac{\int_0^{l_{max}(\hat\theta)} \rho^2 [r(l)] dl(\theta)}{\int_0^{l_{max}(\hat\theta)} \rho_\text{h}^2 [r(l)] dl(\theta)}\,.
\end{eqnarray}
The angular analysis is independent on the particle physic model. In fact, any assumption on the DM mass and emission spectra in  Eq. (\ref{eq:totalflux}) is considered to be the same in both the spike and the halo.

Because of the uncertainties on the background model, we performed a set of different analyses summarized in 
Fig. \ref{HESSIvsII}:
In the first line, we follow the extreme approach of assuming as background the inner extrapolation of the DM-halo profile, so that the angle by angle number of background events in the model depends on the halo profile itself. In this way we investigate the possibility that the background level increases trough the GC with different power laws. The first column corresponds to count the excess events (with respect to the background) from the inner extrapolation of  the DM-halo profiles as measured in different simulations. Normalization to the DM-halo profile itself obviously contains no information in the first case as it goes to one at all radii. For this reason,  Panel (a,1) 
just shows the ON/OFF HESS data. 
In the other two lines of panels, to normalize the ON source, we use an integrated constant value for $d \Phi_{\text{DM-halo}}(\theta)$ associated with two possible normalization of the OFF signal at different angles; this angle is assumed to be either $\theta\approx 0.54^\circ / r\approx80$ pc (where the spatial tail disappears) or $\theta\approx 0.13^\circ / r\approx20$ pc (very close to the extent of the $40\sigma$ signal). The normalization to $\theta \approx 0.54^\circ$ (second line) shows that only the shallowest profiles, GARR-II300 and GARR-I, have an excess consistent with the data (Panel (b,1)). Finally, normalizing to 
$\theta\approx 0.13^\circ$ favors more cusped profiles 
as ERIS and GARR-I300 (Panel (c,1)). The latter line in Fig. \ref{HESSIvsII} is a zoomed view (0-25 pc) with respect to the first two lines (0-100 pc). In the second column, we show the results for the profiles with the DM-spike, as modified by the BH. When normalizing to the DM-halo profile (Panel (a, 2)), the spatial extent of the tail can be barely consistent with the DM-spike associated to more cusp-like halo profiles, such as EAGLE, GARR-I and I-300, that are able to reproduce the spike around 10 pc. However, the count of excess events is not well reproduced. When normalizing to $\theta\approx 0.54^\circ$ (Panel (b,2)) and $\theta\approx 0.13^\circ$ (Panel (c,2)) this results in that only the shallowest inner DM-halo profile, GARR-II300, is consistent with the spatial tail morphology. Finally, in the third column (Panels (a,3), (b,3), (c,3)), we see that the effect of the stars does not change significantly the analyses. This is because such an effect is important at radii inner than those resolved by HESS. Interestingly, our analysis could be useful to constraint different DM-halo profiles with the information given by the spatial resolution of $\gamma$-ray sources.\\

{\bf{Discussion.-}} Strong uncertainties still affect the knowledge of the DM distribution in the inner 100 pc from the GC. It is commonly accepted that a deviation from a merely extrapolation of NFW only-DM profile exists in the inner part. This fact can be related to baryonic effects and to a BH-induced DM-spike. We study this two cases, taking into account also the effect of the stars on the DM-spike. We found that different cases can produce enhancement from $20$ up to more than $10^3$ in the astrophysical factor for DM indirect detection with $\gamma$-rays: the higher value appears to be compatible with what is expected in order to fit the HESS J1745-290 $\gamma$-ray cut-off as TeVDM \cite{Gammaldi}. Moreover, we study the spatial tail of this signal. Different techniques of the background modeling may affect the number counts of the $\gamma$-rays events in excess respect to the OFF background region \cite{BgPSF, Bg}. An extensive analysis is out of the scope of this paper. However, we emphasize the idea that the spatial extent of the tail of such a VHE $\gamma$-ray source could be intrinsically related to the inner (10-100 pc) DM distribution around the BH. In this regard, our analysis suggests that the DM-spike at the GC has already been detected. However, is necessary to well justify the number of ON-source events with respect to the OFF-source background and the PSF deconvolution method. Our analysis with the normalization with constant background measured at $\theta\approx 0.54^\circ$ ($\approx$ HESS field of view) suggests that a shallow underlying DM-halo profile in the innermost regions (as GARR-II300) is consistent with the HESS spatial tail. In this case, the DM-spike could account for an enhancement of only $\lesssim 10\times$ 
in the astrophysical factor with respect to $\langle J \rangle_{\Delta\Omega}^{\rm EVANS}$,
which is much than the requirement of $\sim 10^3$ to fit the HESS $\gamma$-ray spectra as TeVDM. The deficit in the enhancement should be then associated to a different value of the annihilation cross section, such as the one for self annihilating or no-thermal DM. 
On the other hand, cuspy underlying DM-halo profiles in the innermost regions, as GARR-I300 and EAGLE,
lead to boost factors up to $\sim 10^3$, 
while the spatial morphology of such a DM tail appears barely compatible with the HESS resolution, provided that the background scales as the DM-halo density profile. The effect of stars on the DM-spike only slightly lowers the boost factors in all the cases, and is almost imperceptible in the tail morphology at the angles resolved by HESS. Finally, if the underlying DM-halo profile presents a large core as in the case of the MOLL
 simulation, then neither the $J$-factor nor the morphology of the tail could be consistent with the VHE $\gamma$-ray data interpreted as DM annihilation.

{\bf Acknowledgements}
This work has been supported by Beca Iberoam\'erica J\'ovenes Profesores e investigadores by the MINECO (Spain) project FIS2014-52837-P and Consolider-Ingenio MULTIDARK CSD2009-00064, and partially by the H2020 CSA Twinning project No.692194, ÒRBI-T-WINNINGÓ and QGSKY. VG is grateful to the Mainz Institute for Theoretical Physics for its hospitality and its partial support during the completion of this work. AXGM acknowledges the CATEDRA CONACYT program, the UCMEXUS collaborative project, and the UGTO basic science project  878/2016. OV acknowledges CONACyT/Fronteras de la Ciencias 281. The authors acknowledge S. Roca-Fabrega and B. Villasen\~nor for
providing the Garrotxa simulation data and fits, and M. Schaller for
providing the fits to the EAGLE-APOSTLE simulations.

\end{document}